\documentclass[onecolumn]{aastex63}
\usepackage{rotating} 
\usepackage{amsmath,amssymb,bbold,mathrsfs}
\usepackage{subfigure}
\usepackage{bm}
\usepackage{hyperref}

\makeatletter
\let\frontmatter@title@above=\relax
\makeatother

\shorttitle{\ion{Fe}{2} polarization between 250~-~278~nm}
\shortauthors{Afonso Delgado et al.}

\graphicspath{{./}{figures/}}

\begin{document}

\title{The magnetic sensitivity of the (250~–~278 nm) \ion{Fe}{2} polarization spectrum.}

\email{dafonso@iac.es}

\author{David Afonso Delgado}
\affil{Instituto de Astrofísica de Canarias, E-38205 La Laguna, Tenerife, Spain}
\affil{Departamento de Astrof\'isica, Universidad de La Laguna, E-38206, La Laguna, Tenerife, Spain}
\author{Tanaus\'u del Pino Alem\'an}
\affil{Instituto de Astrofísica de Canarias, E-38205 La Laguna, Tenerife, Spain}
\affil{Departamento de Astrof\'isica, Universidad de La Laguna, E-38206, La Laguna, Tenerife, Spain}
\author{Javier Trujillo Bueno}
\affil{Instituto de Astrofísica de Canarias, E-38205 La Laguna, Tenerife, Spain}
\affil{Departamento de Astrof\'isica, Universidad de La Laguna, E-38206, La Laguna, Tenerife, Spain}
\affil{Consejo Superior de Investigaciones Científicas, Spain}

\begin{abstract}
This paper presents a theoretical investigation on the polarization and magnetic 
sensitivity of the near-ultraviolet (near-UV) solar spectral lines of \ion{Fe}{2} between 250 and 278~nm. 
In recent years, UV spectropolarimetry has become 
key to uncover the magnetism of the upper layers of the solar chromosphere. 
The unprecedented data obtained by the CLASP2 
suborbital space experiment across the \ion{Mg}{2} h and k lines around 280~nm
are a clear example of the capabilities of near-UV spectropolarimetry for the magnetic field 
diagnostics throughout the whole solar chromosphere.
Recent works have pointed out the possible complementary 
diagnostic potential of the many \ion{Fe}{2} lines in 
the unexplored spectral region between 250 and 278~nm, 
but no quantitative analysis of the polarization 
and magnetic sensitivity of those spectral lines has been carried out yet. 
To study the polarization signals in these spectral lines, 
we create a comprehensive atomic model including 
all the atomic transitions resulting in strong spectral lines.
We then study the magnetic sensitivity of the linear and circular polarization profiles 
in a semi-empirical model representative of the quiet sun.
We present a selection of \ion{Fe}{2} spectral lines 
with significant linear and circular polarization signals and evaluate 
their diagnostic capabilities by studying their formation heights and magnetic 
sensitivity through the action of the Hanle and Zeeman effects.
We conclude that when combined with the
CLASP2 spectral region these \ion{Fe}{2} lines are of interest for the inference of magnetic fields
throughout the solar chromosphere, up to near the base of the corona.
\end{abstract}



\section{Introduction} \label{sec:intro}

Spectral lines encode information about the physical conditions of the emitting plasma. 
Therefore, the study of their formation in given models of the solar atmosphere 
is very important for reaching the goal of inferring its physical properties 
from the observed spectral line radiation.    
In particular, the linear and circular polarization profiles carry 
information about the magnetic field strength and geometry in the their region of formation.
Even though the solar spectrum is populated by plenty of spectral lines forming everywhere
in the solar atmosphere, it is in the ultraviolet (UV) 
region of the spectrum where we find abundant spectral lines forming in the solar chromosphere
and transition region. Determining the magnetic field in these 
regions of the upper solar atmosphere is of fundamental importance for understanding how the energy 
is transported from the underlying layers and dissipated in the chromosphere and the corona, where 
the explosive events that can impact the heliosphere take place.
For these reasons, ultraviolet spectropolarimetry is increasingly recognized as a 
key diagnostic tool for facilitating new breakthroughs in our empirical understanding of solar and stellar 
magnetic fields\cite[e.g., the review of][]{TrujilloBueno2022}.

The \textit{Interface Region Imaging Spectrograph} \cite[IRIS,][]{dePontieu2014} mission is one of the main
examples of the relevance of UV spectroscopy for investigating 
the upper layers of the solar atmosphere. IRIS has been in operation since 2013
measuring the intensity profiles
in several windows of the UV solar spectrum, including spectral lines such as
\ion{Mg}{2} h and k (279.6 and 280.3~nm), \ion{C}{2} (133.4/133.5~nm), or \ion{Si}{4} (139.4/140.3~nm).
The research exploiting the data provided by the IRIS 
mission has significantly improved our knowledge about 
the solar chromosphere \citep[see][]{dePontieu2021}.

Concerning UV spectropolarimetry, the most recent relevant advances 
are the \textit{Chromospheric Lyman-Alpha SpectroPolarimeter}
\cite[CLASP,][]{CLASP}, which in 2015 observed
the intensity 
and linear polarization of the \ion{H}{1} Ly~$\alpha$ line \citep{Kano2017,TrujilloBueno2018}, and the 
\textit{Chromospheric LAyer SpectroPolarimeter} \cite[CLASP2,][]{CLASP2}, which
in 2019 acquired unprecedented spectropolarimetric data of the 279.2~-~280.7~nm 
near-UV spectral region that contains the
\ion{Mg}{2} h and k lines and several other lines of great diagnostic interest 
\citep{Ishikawa2021a,Rachmeler2022}. 
These suborbital space missions were motivated by a series of  
theoretical investigations \citep[see][]{TrujilloBueno2011,TrujilloBueno2012,Belluzzi2012a,Belluzzi2012,
AlsinaBallester2016,DelPinoAleman2016}, 
which predicted that the polarization 
signals are measurable and magnetically sensitive and showed the importance of 
the effects of
partial frequency redistribution (PRD),
J-state interference, and the joint action of the Hanle, Zeeman, and magneto-optical effects
on the linear polarization signals produced by scattering processes. 
Moreover, the analysis of the circular polarization profiles 
measured by CLASP2 in the resonance lines of \ion{Mg}{2} and \ion{Mn}{1}
allowed for the first mapping
of the longitudinal component of the magnetic field from the photosphere to 
the upper chromosphere, just below the transition region \citep{Ishikawa2021a,Li2023}.

Despite all these novel advances, there are still many unexplored regions in the UV spectrum
of the Sun. The polarization signals of the near-UV \ion{Fe}{2} spectral lines 
that are located blueward of the CLASP2 spectral region
have been proposed
as potentially useful for facilitating the exploration of chromospheric magnetism \citep{Judge2021}. 
In particular, in the 250~-~278~nm spectral window we can find many 
lines belonging to this ion, including a
 strong resonant multiplet at 260~nm. Due to the lack of spectroscopic or spectropolarimetric 
 solar observations in this spectral region, previous works found in
 the literature are mainly focused on the study of the intensity spectrum
 of these spectral lines in other stars \citep{Judge1991,Judge1992}
 or of other \ion{Fe}{2} spectral lines located in observed spectral regions,
 such as the small \ion{Fe}{2} emission line in the red wing of the \ion{Ca}{2} line at 369.94~nm 
 \citep{Cram1980,Watanabe1986}. Moreover, as far as we know, there are not theoretical
 investigations about the polarization signals of these spectral lines or about their magnetic
 sensitivity.
 
In this work we present the first detailed theoretical study 
on the polarization and magnetic sensitivity of the \ion{Fe}{2} 
lines that are located in the 250~-~278~nm near-UV spectral region,
as well as their suitability to infer magnetic fields in the solar atmosphere. In Sec.~\ref{sec:problem} we introduce the methods and tools used 
in our radiative transfer modeling. The atomic model used in the calculation is
described in Sec.~\ref{sec:atomicModel}. In Sec.~\ref{sec:Results} we show the
results of our modeling, including a selection of the most interesting spectral lines, following certain criteria,
as well as an analysis of their magnetic sensitivity and suitability to infer magnetic fields in
the solar atmosphere. Finally,
present our conclusions in Sec.~\ref{sec:conclusions}.


\section{Formulation of the problem} \label{sec:problem}

To study the formation and magnetic sensitivity of the \ion{Fe}{2} lines in the spectral window between 250 and 278~nm
we solve the problem of generation and transfer of polarized radiation in optically-thick magnetized plasmas out of local thermodynamic equilibrium (NLTE).
This problem is both non-local and non-linear, requiring the simultaneous solution of the atomic 
excitation (populations of the atomic substates and quantum coherence between them)
 described by the statistical equilibrium equations (SEE) and of the
radiation propagation throughout the atmosphere governed by the radiative transfer (RT) equations.
We solve this problem by applying the HanleRT code \citep{DelPinoAleman2016,delPinoAleman2020} to a 1D plane-parallel model of the solar atmosphere.
We have chosen the semi-empirical C model of the quiet Sun described in \citet[][hereafter FAL-C]{FALC}.

Even though PRD effects and quantum interference can have an impact
on the linear scattering polarization profiles \citep[e.g.,][]{Casinietal2017c}, we neglect both for
the following reasons: first, these effects are particularly relevant in the far wings of these spectral lines
and, as we discuss in more detail in Sec.~\ref{sec:atomicModel}, in the near-UV region of the solar
spectrum the effect of spectral line blanketing is
notably strong, masking the impact of these effects and making the use of the line wings for atmospheric
diagnostic extremely difficult. Second, including these physical ingredients has a significant computational
cost, especially when considering the size of the atomic 
model needed to study all the spectral lines considered
in this work. For these reasons, the calculations 
shown in this paper assume complete frequency 
redistribution (CRD) and neglect quantum interference between the substates pertaining to different 
atomic levels, which are often good approximations to model the spectral line core.


\newpage

\section{Atomic model} \label{sec:atomicModel}

\begin{figure*}[ht!]
\plotone{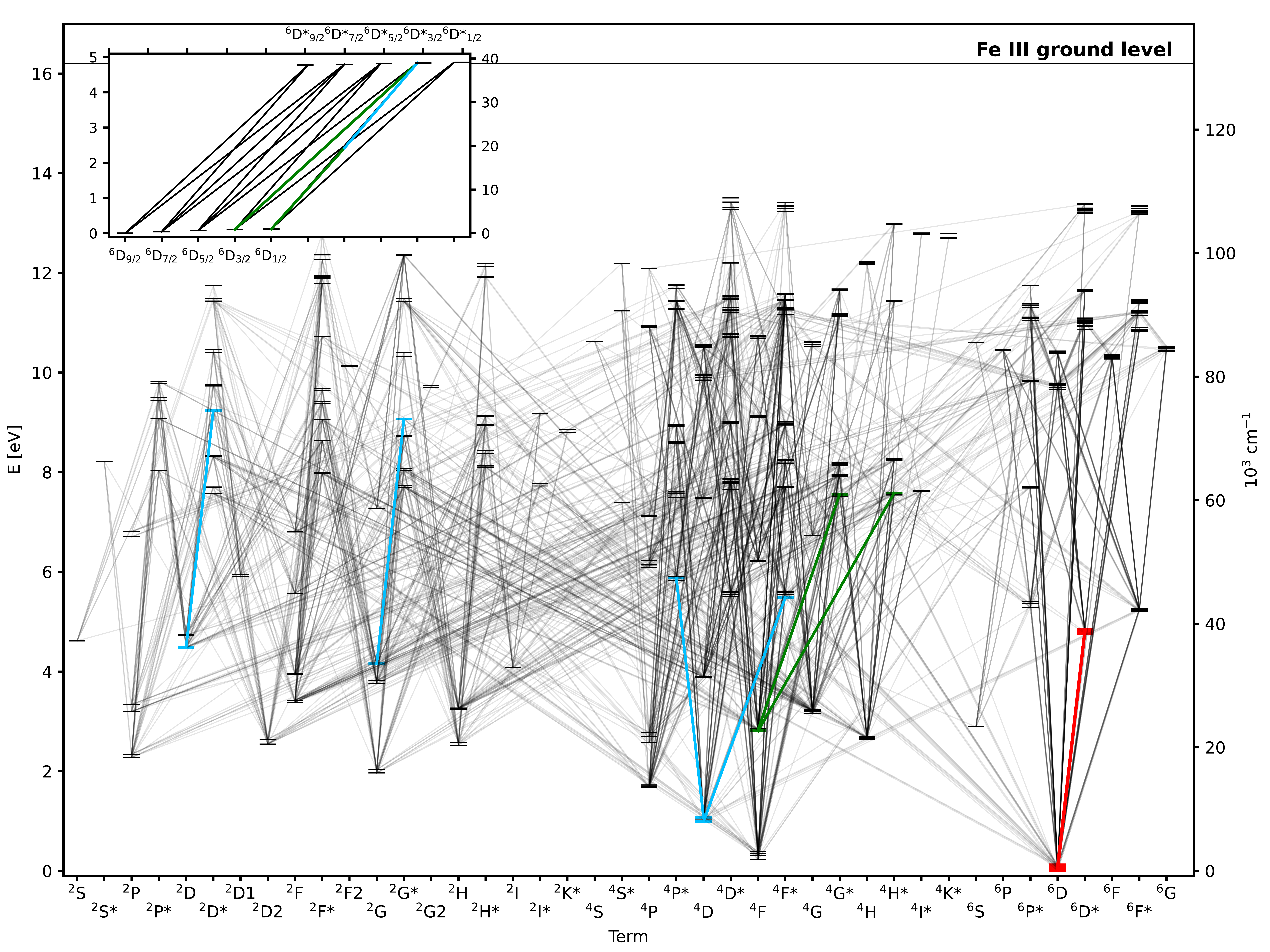}
\caption{Grotrian diagram of the \ion{Fe}{2}
atom considered in this work. The red line 
indicates the resonant multiplet at 260~nm, 
shown with further detail in the Grotrian diagram
in the inset at the upper-left. The blue (green) lines
indicate the transitions
 shown in Fig.~\ref{fig:QMF} (Fig.~\ref{fig:wfaFe}). \label{fig:grotrian}}
\end{figure*}

As far as we know, there are not previous studies on the formation of the \ion{Fe}{2}
polarization profiles in the spectral window between 250 and 278~nm
and, therefore, on the atomic model necessary to reliably
model the polarization of all these spectral lines.
Consequently, the first step is to build a suitable atomic model including all the
atomic transitions that could produce (or impact) any interesting polarization signal
in the spectral region of interest.

In the solar chromosphere the iron atoms are mostly singly ionized, and, consequently, \ion{Fe}{2} 
is the dominant ion \citep[e.g.,][]{Shchukina2001}.
Our atomic model thus includes just the \ion{Fe}{2} levels and the ground level of \ion{Fe}{3}.
We take the atomic level energies and the Einstein coefficients for spontaneous emission from the
NIST database \citep{NISTbib}.
We approximate the photoionization cross sections as hydrogenic \citep{MihalasBook}. The
rates of ionizing inelastic collisions with electrons are
calculated using the approximation
of \cite{AllenBook1963}.
The rate of excitation due to inelastic collisions with electrons are calculated following \cite{vanRegemorter1963} 
for pairs of levels fulfilling the electric dipole selection rules, and
following \cite{Bely1970} for pairs of levels that do not fulfill those selection rules.
The rate of depolarizing collisions with neutral hydrogen atoms are calculated assuming
the Van der Waals potential \citep{LambTerHaar1971,LL04}. We neglect inelastic collisions with
hydrogen atoms.

In our atomic model we only include those transitions with a spontaneous emission probability above
a certain threshold ($A_{ul} \geq 10^6$~s$^{-1}$), which can produce a significant (in terms of its depths in the emergent spectrum) spectral line. Moreover,
once we have
solved the self-consistent RT problem with the whole atomic model we then remove one third of the levels, the least
populated ones, as well as those not radiatively connected with any other atomic level. Using the reduced atomic model results in relative differences of no more than 5\% at the center of the emergent intensity profiles for the spectral lines in the 250 -- 280~nm range.

Our atomic model to study the formation and polarization of the \ion{Fe}{2} spectral lines
between 250 and 278~nm
has 453 \ion{Fe}{2} atomic levels (and the ground level of \ion{Fe}{3})
and
2225 radiative transitions.
The model is thus complete enough to ensure a reliable calculation of the population balance of the levels involved in
the formation of any of the spectral lines in the range between 250 and 280~nm.

In Fig.~\ref{fig:grotrian}
we show the Grotrian diagram for the
described atomic model, including all the radiative transitions.
The red line in the diagram indicates
the resonant multiplet that produces a series of intense spectral lines around 260~nm.
A detailed Grotrian diagram of this multiplet is shown in the  inset at the top-left corner of Fig.~\ref{fig:grotrian}.
The blue and green lines highlight some spectral lines analyzed in detail
in Secs.~\ref{sec:StkQ} and \ref{sec:StkV}.


\begin{figure*}[ht!]
\plotone{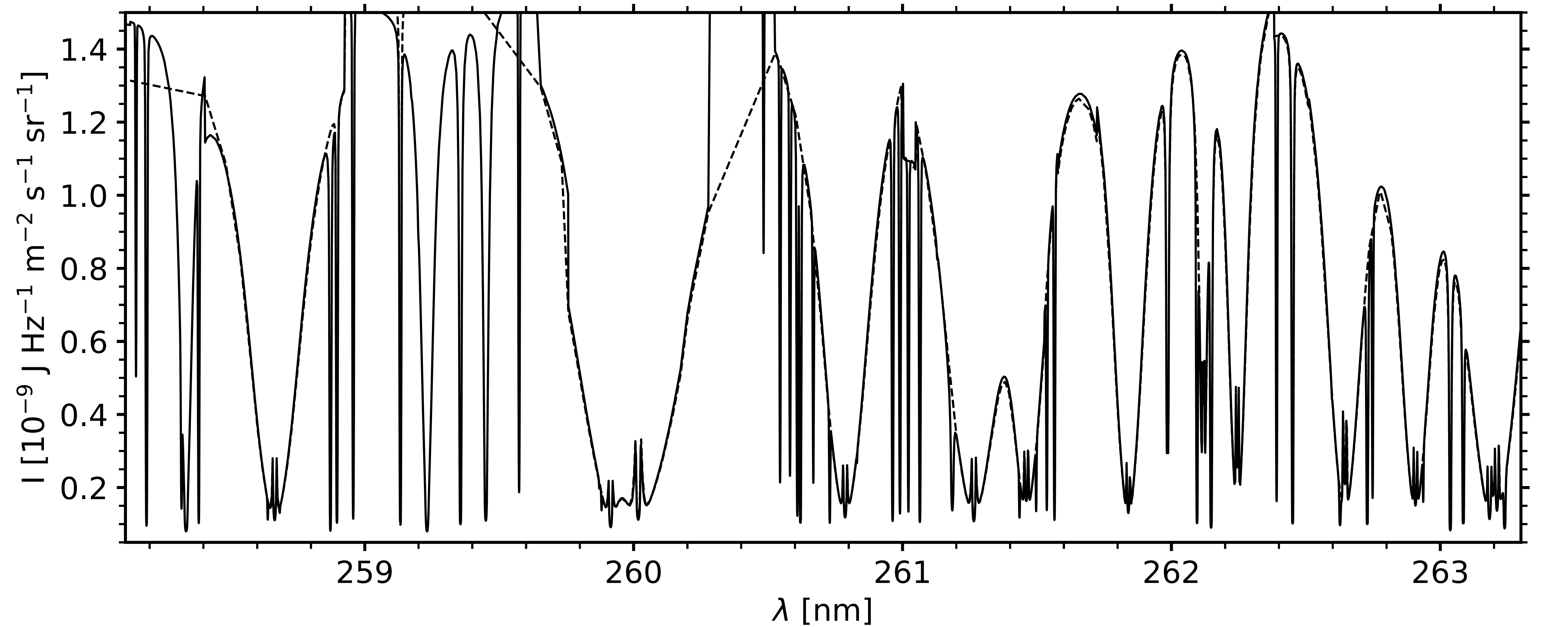}
\caption{Intensity profile in the spectral region of the
\ion{Fe}{2} resonant multiplet around 260~nm. Solid line: atomic model with 453 levels.
 Dashed line: atomic model with 90 levels. \label{fig:ModelComparation}}
\end{figure*}

The rates of inelastic collisions are critical for the population balance of the \ion{Fe}{2} atomic
levels \citep{Judge1992}. To build our atomic model we use very rough approximations to these rates, what
can have a significant effect on the emergent Stokes profiles. 
In order to study how the inelastic collisions
affect the emergent profiles, we create a reduced \ion{Fe}{2} atomic model with 90 levels (including
the \ion{Fe}{3} ground level) and 113 radiative transitions, which can correctly model
the emergent intensity profiles of the resonant multiplet around 260~nm; 
that is, it is able to reproduce the same emergent intensity profiles for these strong resonant lines
(Fig.~\ref{fig:ModelComparation}).
Using the reduced model, we perform different 
numerical experiments in which we multiply the inelastic collisional 
rates between pairs of atomic levels fulfilling (not fulfilling) 
the electric dipole selection rules with a factor $\Omega_A$ ($\Omega_F$).

\begin{figure*}[ht!]
\plotone{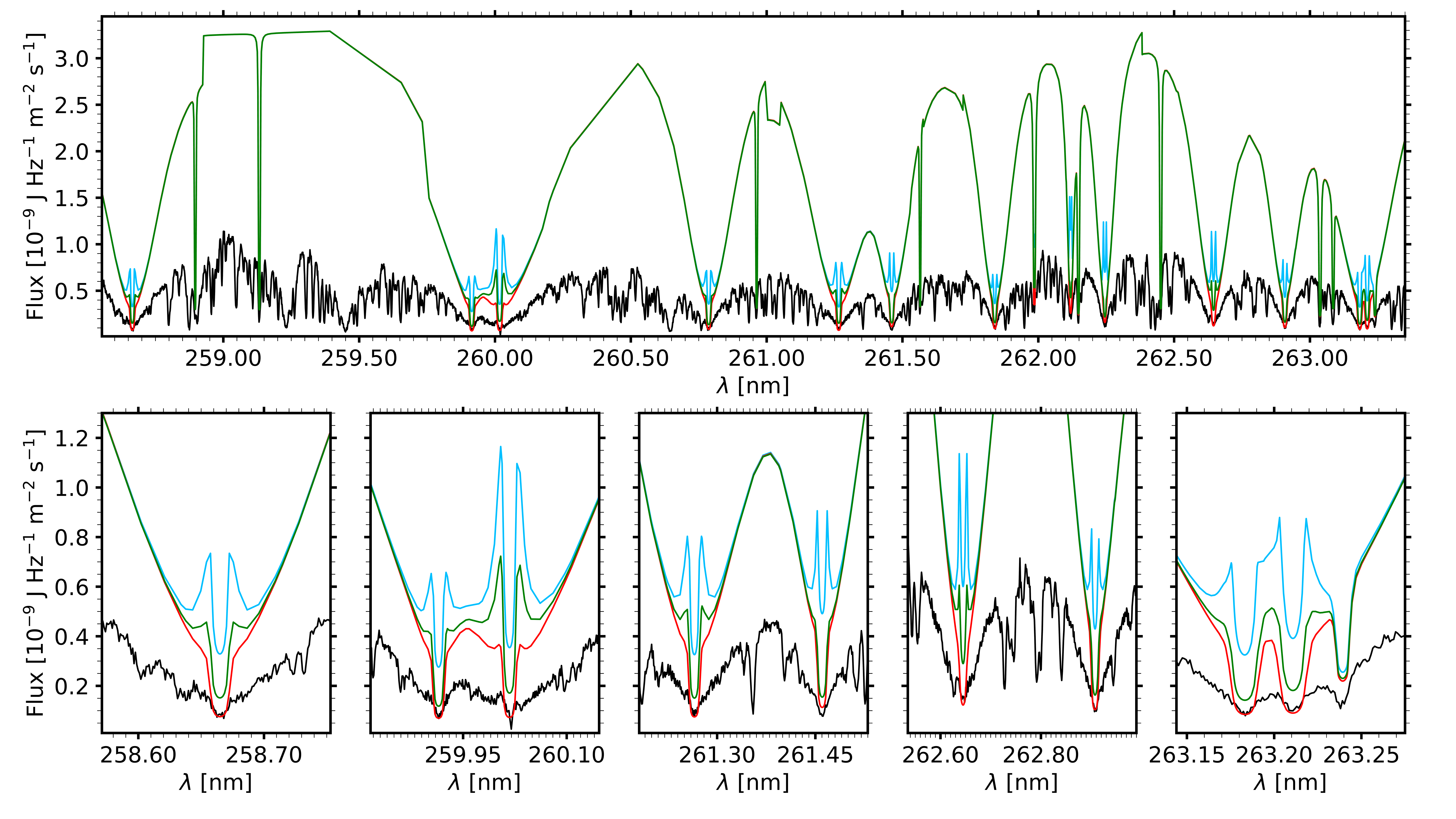}
\caption{Emergent intensity flux profile in the spectral region of the \ion{Fe}{2} resonant
multiplet at 260~nm. The black
curve shows the flux of $\alpha$ Cen A (a 
 G2V star) measured with the STIS instrument of the HST. Each
colored curve shows the 
flux profile calculated with different inelastic collisional rates, resulting by scaling the original
atomic model values (see Sec.~\ref{sec:atomicModel}) by $\Omega_{A}$ ($\Omega_{F}$) for pairs of levels that
fulfill (do not fulfill) the electric dipole selection rules:
 $\Omega_{F} = \Omega_{A}$ = 1 (blue), $\Omega_{F} = 0.1$ and $\Omega_{A} = 1$ (green), and $\Omega_{F} = \Omega_{A} = 0.1$ (red). The bottom row shows
in more detail the core of the
strongest transitions. \label{fig:IneCol}}

\end{figure*}

Due to the lack of solar spectroscopic observations, it is not possible to compare our calculations with the solar spectrum at these wavelengths.
We have thus opted for a comparison with the
flux of a solar analog star, $\alpha$ Cen A, a
very close G2V star \citep{Torres2006}, the same spectral type than the Sun.
We compare the intensity flux profile of $\alpha$ Cen A\footnote{ This spectrum is available in The Advanced Spectral Library Project (ASTRAL, \url{https://archive.stsci.edu/prepds/astral/}). For further details see \cite{Ayres2013, Ayre2010}}  observed with the \emph{Space Telescope Imaging Spectrograph} \cite[STIS,][]{Woodgate1998} on board of the \emph{Hubble Space Telescope} (HST) with calculations in the FAL-C model for different values of $\Omega_A$ and $\Omega_F$ (Fig.~\ref{fig:IneCol}).
The first evident difference between
the theoretical and observed fluxes is the significantly diminished continuum intensity in the observation, showing considerably wider lines. 
The large amount of spectral lines in the UV spectrum, 
contributing to the opacity, produce the so called UV blanketing, 
which consistently reduces the intensity flux across the spectral region,
giving the impression of a much lower continuum and of wider spectral lines.
Consequently, we focus our analysis to the core and near wings of the \ion{Fe}{2} spectral lines.

With the inelastic collisions given by the above-mentioned approximations 
($\Omega_A=\Omega_F=1.0$; 
blue curve) the flux in the line core is significantly larger than the observed flux, with significant
emission features in the near wings which are absent in the observation.
We find the best agreement when all the inelastic collisional rates are reduced by a factor of ten
($\Omega_A=\Omega_F=0.1$; red curve); the intensity flux at the line cores becomes remarkably similar to the 
observed flux and the near wing emission features completely vanish (see bottom row of Fig.~\ref{fig:IneCol}).
Our atomic model is thus able to adequately
reproduce the intensity flux of the most intense \ion{Fe}{2} atomic lines in the near-UV
spectrum of a solar-like star.


\section{Results} \label{sec:Results}

We use the atomic model described in Sec.~\ref{sec:atomicModel} to
model the Stokes profiles of the \ion{Fe}{2} atomic lines in the UV spectral
window between 250 and 278~nm.
In particular, we study their linear and circular polarization
profiles and their capability to uncover the magnetic fields of the solar atmosphere.
For this purpose, we perform RT calculations
including scattering polarization and the Hanle and Zeeman effects 
assuming CRD. We study the fractional linear polarization signals
for a line-of-sight (LOS) near the solar limb
($\mu=\cos{\theta}=0.1$, with $\theta$ the heliocentric angle).
We then impose different magnetic fields to investigate the magnetic sensitivity of their
linear polarization.
Finally, we study the formation of Zeeman-induced circular polarization signals
for the disk center LOS.
We then study the suitability of the weak-field approximation (WFA) to infer the longitudinal
component of the magnetic field ($B_L$) from the circular polarization.

\subsection{Linear Polarization Signals.} \label{sec:StkQ}

\begin{figure*}[ht!]
\epsscale{1.2}
\plotone{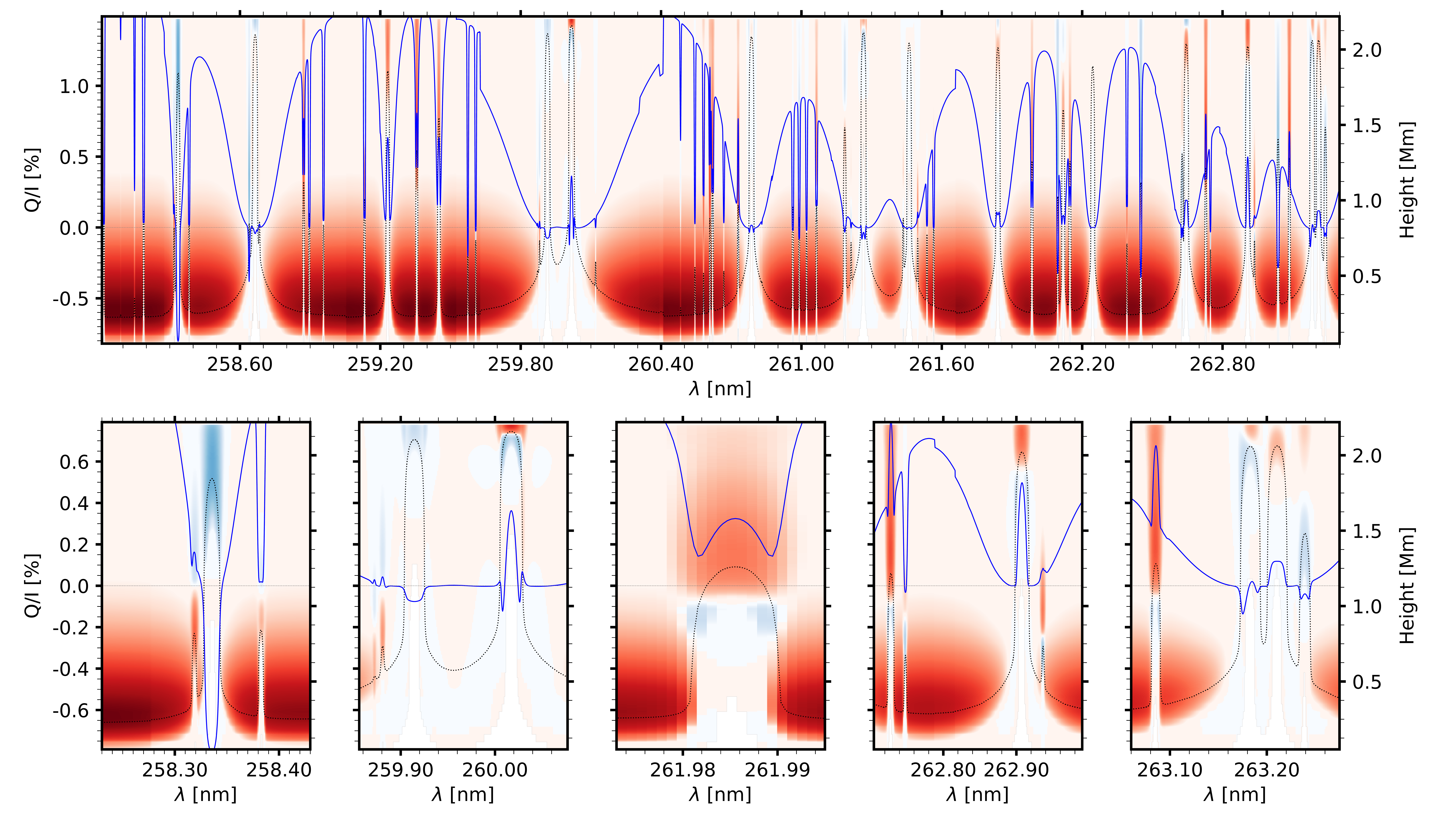}
\caption{
Fractional linear polarization profile $Q/I$ (solid blue curve, left plot axes) and height where the
optical depth is equal to one (dotted black curve, right plot axes)
in the spectral region of the \ion{Fe}{2} resonant multiplet around 260 nm.
The colored background
shows the logarithm of the normalized contribution function for Stokes $Q$
(with the red and blue colors indicating positive and negative contributions, respectively).
Details around some of the lines with the strongest $Q/I$ signals are shown in the bottom row.
\label{fig:Q260}}
\end{figure*}

Fig.~\ref{fig:Q260} shows the fractional linear polarization $Q/I$ profile in the unmagnetized FAL-C model atmosphere for the spectral region
of the \ion{Fe}{2} resonant multiplet around 260~nm for the LOS at $\mu=0.1$ (see blue curve in Fig.~\ref{fig:Q260}).
While some of the strongest lines, in terms of their absorption features in the intensity profile,
hardly show any significant linear polarization signal (see
lines at 259.914, 261.265, or 261.840 nm)
others exhibit relatively large signals
shown in detail in the
bottom panels of Fig.~\ref{fig:Q260}.
The colored background in Fig.~\ref{fig:Q260} shows the logarithm of the normalized contribution function for Stokes $Q$,
which gives a measure of how much each height layer in the model atmosphere contributes, at each
wavelength, to the linear polarization of the emergent radiation. The red
and blue colors indicate positive or negative contributions, respectively.
We can conclude, from the contribution function, that the linear polarization at the line core of the
lines shown in the figure is formed between the middle and upper chromosphere (above $\approx 1$~Mm in the FAL-C model).

\begin{figure*}[ht!]
\epsscale{1.2}
\plotone{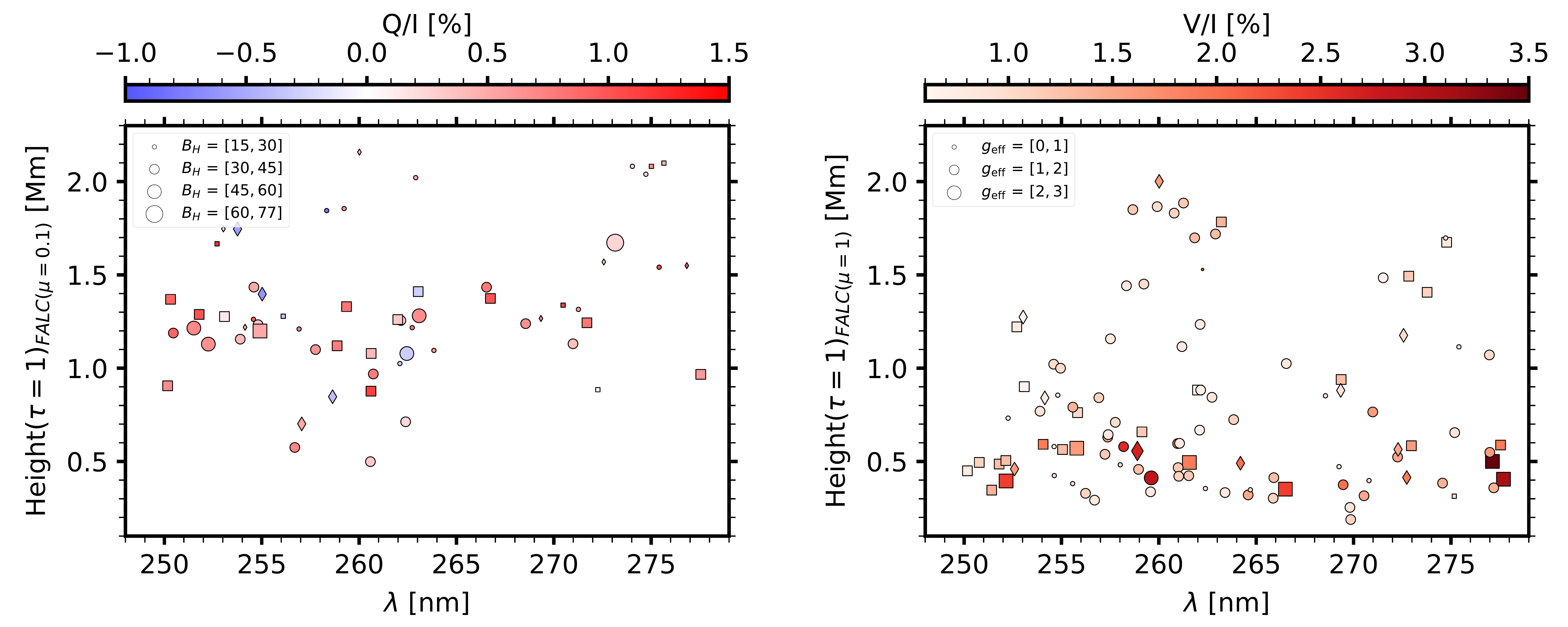}
\caption{
The left (right) panel shows a selection of the \ion{Fe}{2} 
spectral lines with larger fractional linear (circular) polarization $Q/I$ ($V/I$) signals for the LOS
at $\mu=0.1$ ($\mu=1$). The size of the markers indicates the critical Hanle field $B_{H}$
(the effective Land\'e factor $g_{\rm eff}$) for each spectral line. The color shows the
amplitude of the $Q/I$ ($V/I$,
for a longitudinal magnetic field of 50~G)
signals for each spectral line. The markers's shape
indicate the presence and
significance of blends with
other spectral lines: circle = no blends, square = moderate blend, and diamond = severe blend
 (see text for further details).
\label{fig:linesSel}}
\end{figure*}

We have selected the spectral lines 
with zero-field scattering polarization signals $Q/I>0.2$~\% at the line core, and we have  
analyzed their magnetic sensitivity in order to determine which ones are suitable 
for the diagnostics of chromospheric magnetic fields.
In the left panel of Fig.~\ref{fig:linesSel}
we show the height where the optical depth is unity at the line center (which gives an idea of the region 
of formation) for the selected lines, as well as their Hanle critical field ($B_H$), their line center zero-field $Q/I$ signal, and information about the blends with neighbor lines. The Hanle critical field
is the magnetic field strength for which the Zeeman splitting equals the natural width of
the line \citep[e.g.,][]{TrujilloBueno2001}:

\begin{equation}
    B_H = \frac{1.137\cdot10^{-7}}{t_{\rm life}g} ,
\end{equation}
with $t_{\rm life}$ and $g$
the radiative lifetime and Landé factor of the line level under consideration, respectively. Typically, 
a spectral line is sensitive, via the Hanle effect, to magnetic fields with strengths between
approximately $0.2B_H$ and $5B_H$.

To characterize the blends with other spectral lines we compare the theoretical and observed fluxes
from Fig.~\ref{fig:ModelComparation} with a synthetic profile including all the transitions from the Kurucz
database in LTE \citep{KuruczDB}, similarly to the approach of \cite{Judge2021}.
Through visual inspection, we look for any spectral line close enough 
to the selected \ion{Fe}{2}  line to 
affect their profiles, and we classify them in three categories: no nearby blends (circles in
Fig.~\ref{fig:linesSel}), a moderate blend in the wings, far enough from the line core (squares), 
or a significant blend close to the line core (diamonds). Many of the selected lines show a nearby blend but,
in these cases, the line core (line region sensitive to the Hanle effect) should not be affected
significantly.

Most of the spectral lines in our selection form in the middle chromosphere (around $1.25$~Mm in the FAL-C semi-empirical model). A
relatively small number of lines form in the lower chromosphere and near the FAL-C 
temperature minimum
(around $0.5$~Mm). Moreover, several stronger lines form in the upper chromosphere (above around $1.8$~Mm).
These latter transitions belong to
a resonant multiplet (260.017 and 262.907~nm) and
the metastable multiplets $a^4D$ - $z^4P^{\circ}$ (258.336 and 259.231~nm), $a^4D$ - $z^4D^{\circ}$ 
(274.035~nm), and $a^4D$ - $z^4F^{\circ}$ (274.730, 275.013 and 275.658~nm).

Most of
the lines in our selection have critical Hanle fields below 50~G
and, for those lines forming in the upper
chromosphere, we always
find $B_H<30$~G. Consequently,
these spectral lines are thus expected to be sensitive to magnetic fields with strengths
considerably smaller than 100~G and,
for those lines forming in the upper chromosphere, to magnetic fields around 20~G or less.
These magnetic field strengths are of the order we expect for the chromospheric magnetic field in
the quiet Sun and, therefore, these spectral lines seem to be good candidates to study the quiet Sun
magnetism in the upper chromosphere.

\begin{figure*}[ht!]
\epsscale{1.2}
\plotone{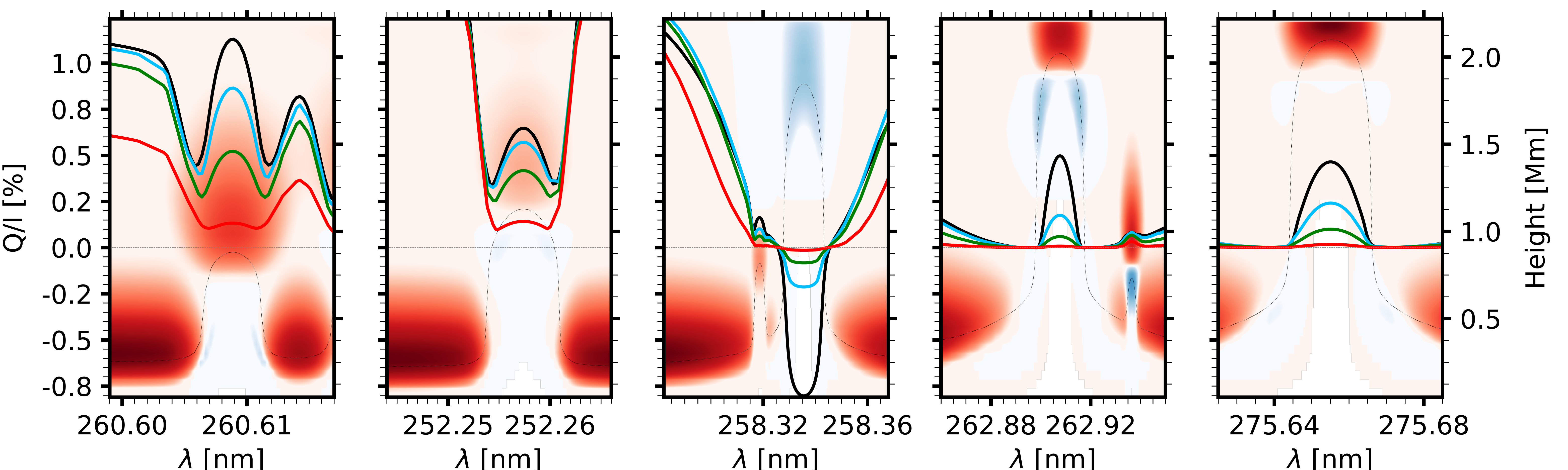}
\caption{
Like in Fig.~\ref{fig:Q260}, but for the five selected spectral lines
(see Tab.~\ref{tab:hanle}) for the LOS at $\mu=0.1$.
The color of the curve indicates the strength of a horizontal 
($\theta_{\rm B}=90^\circ$, $\phi_{\rm B}=0$) and
homogeneous magnetic field in the FAL-C model: 0 (black), 10 (blue), 20 (green), and 50~G (red).
\label{fig:QMF}}
\end{figure*}

To actually study the magnetic sensitivity of these spectral lines, we have selected
five specific cases, choosing lines forming in different layers of the solar atmosphere 
and with different Hanle
critical fields. Table~\ref{tab:hanle} summarizes the atomic data of the five chosen transitions, also
highlighted in blue in the Grotrian diagram of Fig.~\ref{fig:grotrian}.
In the FAL-C model we impose a horizontal magnetic field 
($\theta_{\rm B}=90^\circ$, $\phi_{\rm B}=0$, with $\theta_{\rm B}$
and $\phi_{\rm B}$ the polar and azimuthal angles 
of the magnetic field vector with respect to the local vertical)
with different strengths, and we compare the calculated  
$Q/I$ profiles of the radiation emerging at the $\mu=0.1$ 
LOS (see Fig.~\ref{fig:QMF}): 0 (black), 10 (blue), 20 (green) and 50~G (red).

\begin{deluxetable*}{ccccc}\label{tab:hanle}
\tablecaption{
Wavelength, level configuration and energy, Einstein coefficient for spontaneous emission,
and critical Hanle magnetic field of the upper level for the spectral lines studied in
detail in Sec.~\ref{sec:StkQ}.}
\tablehead{
\colhead{$\lambda$ [nm]} & \colhead{Transition} & 
\colhead{Energies [cm$^{-1}$]} & \colhead{A$_{ul}$ [$10^8$ s$^{-1}$]} & \colhead{B$_{H}^u$ [G]} }

\startdata
260.609 &  b$^2$D$_{3/2}$ - x$^2$D$^{\rm o}_{3/2}$ &          36 126 - 74 498  & $2.00$ & $\sim$40 \\
252.257 &  c$^2$G$_{7/2}$ - w$^2$G$^{\rm o}_{7/2}$   &        33 501 - 73 143  & $2.60$ & $\sim$57 \\
258.336 &  a$^4$D$_{3/2}$ - z$^4$P$^{\rm o}_{3/2}$   &   $\:$  8 680 - 47 389  & $0.88$ & $\sim$20 \\
262.908 &  a$^6$D$_{1/2}$ - z$^6$D$^{\rm o}_{3/2}$ &     $\quad$ 977 - 39 013  & $0.87$ & $\sim$16 \\
275.655 &  a$^4$D$_{7/2}$ - z$^4$F$^{\rm o}_{9/2}$ &      $\:$ 7 955 - 44 232  & $2.20$ & $\sim$22 \\
\enddata
\end{deluxetable*}

The background color in Fig.~\ref{fig:QMF} shows the logarithm of the normalized contribution function as in Fig.~\ref{fig:Q260},
what confirms that the linear polarization in this selection of lines form at different heights
ranging from a few hundred kilometers above the temperature minimum ($>0.5$~Mm) and the upper
chromosphere ($\gtrapprox1.7$~Mm).
For the two metastable transitions at 260.608 and 252.257~nm (first and second panels in
Fig.~\ref{fig:QMF}) most of the contribution is localized in a relatively extense region in the middle
chromosphere of the model. These two transition have $B_H=40$ and 50~G, respectively and, consequently,
a magnetic field of just 10~G already induces a significant depolarization of the line core
linear polarization. A magnetic field of 50~G already reduces the fractional linear polarization
by  more than a factor ten.

Contrarily to the rest of the selected transitions,
the spectral line at 258.336~nm shows a strong negative fractional linear polarization
signal (radial polarization; see third panel of Fig.\ref{fig:QMF})
of about -0.8~\%.
 This line forms in a extense region of the upper chromosphere of the model and it has $B_{H}\approx20$~G.
Consequently, a 10~G magnetic field already more than halved the amplitude of the linear polarization
and it is dramatically depolarized by a 50~G magnetic field.

Finally, regarding the spectral lines
at 262.908 and 275.655~nm, their line core linear polarization forms in a narrow area in the top of the FAL-C model's
chromosphere (between 1.9 and 2.2~Mm and between 2.0 and 2.2~Mm, respectively).
Their critical Hanle fields are also relatively small (16 and 22~G, respectively), as evident
from the strong depolarization induced by a magnetic field of $20$~G (see fourth and fifth panels of
Fig.~\ref{fig:QMF}). Nevertheless, these lines are sensitive to magnetic fields with strengths of a few
gauss, of the order of the typical magnetic field strengths expected in the upper chromosphere of
the quiet Sun. Moreover,
note that the chosen magnetic field geometry, namely horizontal with respect to the local vertical, 
is such that it maximizes the depolarization of the fractional linear polarization $Q/I$ profiles.


\subsection{Circular Polarization Signals.} \label{sec:StkV}

We now study the circular polarization profiles of the \ion{Fe}{2} UV lines and their
suitability to infer the longitudinal component of the magnetic field via the WFA.

For this purpose, we
calculate the fractional circular polarization profiles $V/I$
for a LOS at the disk center ($\mu=1$), imposing in the FAL-C 
model a uniform vertical magnetic field of 50~G.
We show in the right panel of Fig.~\ref{fig:linesSel} those lines with a circular polarization amplitude 
above 0.5\%, with the color of the marker indicating such amplitude and its size indicating the effective
Land\'e factor ($g_{\rm eff}$) calculated assuming L-S coupling.
Although the lobes of the circular polarization profiles typically sample the atmosphere at
regions significantly below those sampled by the line cores, 
we will use the height at which the line-center optical depth is unity to 
indicate the approximate height at which these profiles are formed.

A total of 103 spectral lines show circular polarization signals
larger than 0.5~\%. Most of these lines
form between the upper photosphere and lower chromosphere.
A considerable amount of \ion{Fe}{2} lines 
in the selection form in the middle and upper chromosphere, e.g.,
the transitions of the resonant multiplet $a^6D$ - $z^6D^{\circ}$.

\begin{figure*}[ht!]
\epsscale{0.8}
\plotone{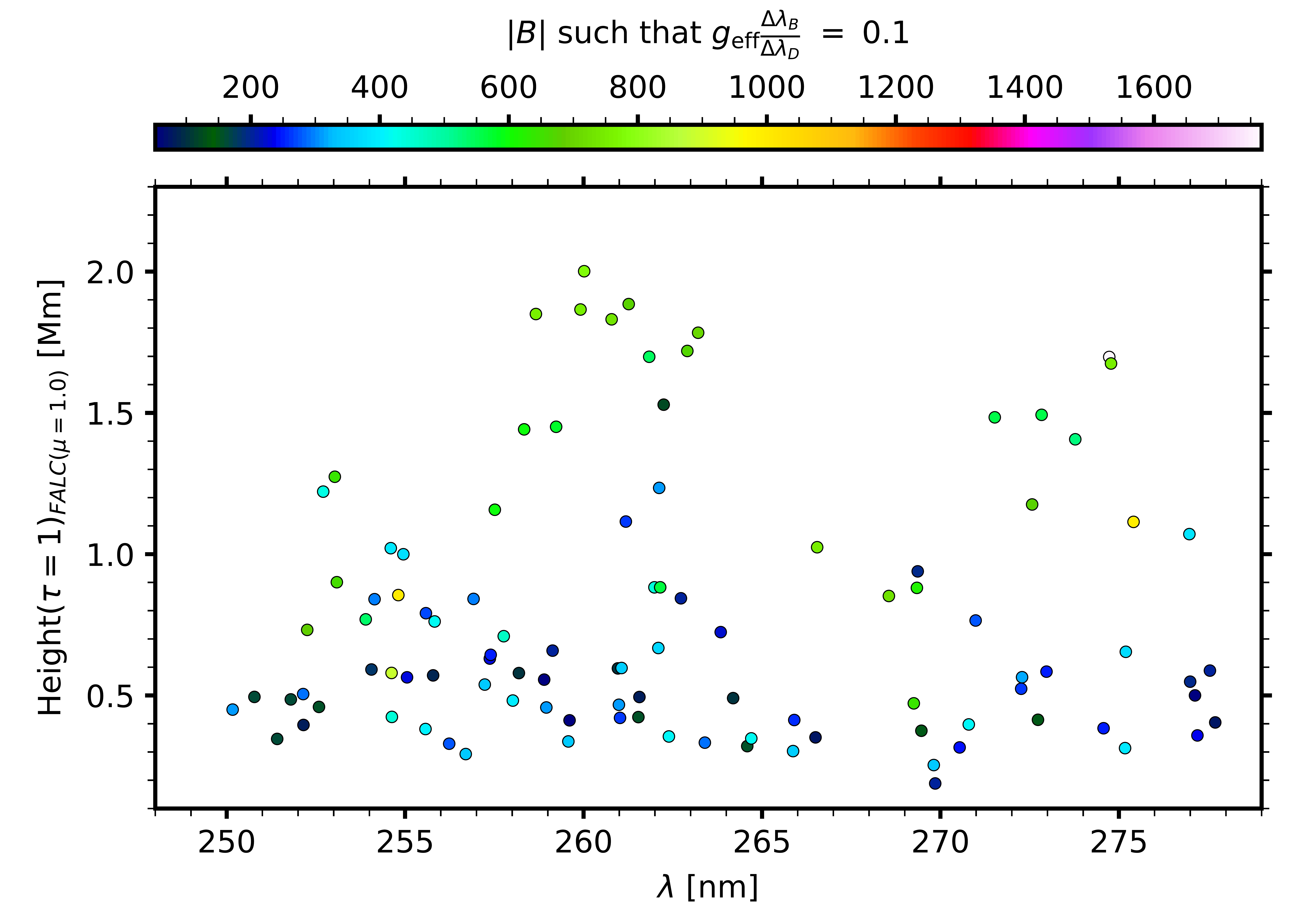}
\caption{Magnetic field strength
for each selected spectral line (see right panel of Fig.~\ref{fig:linesSel}) such that
$g_{\rm eff}\frac{\Delta\lambda_B}{\Delta\lambda_D}=0.1$.
$\Delta\lambda_D$
is calculated 
for the temperature and microturbulent velocity
in the model atmosphere at the height where the optical depth at the line core is equal to one.
This gives an idea of the magnetic field up to which the weak-field condition is fulfilled for
each spectral line.
\label{fig:wfaLim}}
\end{figure*}

Given the computational demands for solving the full NLTE RT problem, it is ideal to have
inference methods that allow for a quick determination of the magnetic field without the need of
carrying out the NLTE modeling of the observations.
The WFA, if suitable, allows for the very quick determination of the longitudinal component
of the magnetic field \cite[e.g., for the \ion{Mg}{2} h and k lines, see][]{Afonso2023}.
For this approximation to be suitable, the line's Doppler width
($\Delta\lambda_D$) must be much larger than the magnetic Zeeman splitting between the atomic
magnetic sublevels ($\Delta\lambda_B$),

\begin{equation}
g_{\rm eff}\frac{\Delta\lambda_B}{\Delta\lambda_D} << 1,
\end{equation}
where $\Delta\lambda_B = 4.6686\cdot10^{-13}B\lambda_0^2$ (with $B$ in Gauss and $\lambda_0$ in \AA) and $\Delta
\lambda_D = \frac{\lambda_{0}}{c} \sqrt{\frac{2kT}{m} + v_m^2}$ (with $T$ and $v_m$
the temperature and turbulent velocity in the formation region of the line, with $c$ the speed of light, 
$k$ the Boltzmann constant, and $m$ the mass of the atom). 
In this way, we can roughly expect the WFA to be valid when the magnetic field strength is below the value 
which makes $g_{\rm eff}\frac{\Delta\lambda_B}{\Delta\lambda_D} = 0.1$.
Fig.~\ref{fig:wfaLim} shows the maximum magnetic field strength for which this condition is fulfilled.
The included lines are those selected in the right panel of Fig.~\ref{fig:linesSel}. 

For those spectral lines whose line core forms between the middle and upper chromosphere 
the WFA is valid, through the chosen criteria, for any magnetic field strength below
500~G (except for the one spectral line at 262.245 nm). We can thus be confident that the WFA necessary condition is met both in quiet sun regions and in the weaker manifestations of active regions, such as plages.

For spectral lines originating between the upper photosphere and lower chromosphere we can distinguish two cases: spectral lines in which the WFA applicability condition is met for $B_L<=200~G$, and thus only applicable in quiet Sun regions, and spectral lines for which the applicability condition is $B_L<=600~G$, and thus it can also be applied in relatively weak active regions such as plages.

\begin{deluxetable*}{ccccc} \label{tab:zeeman}
\tablecaption{
Wavelength, level configuration  and energy, Einstein coefficient for spontaneous emission,
and effective Land\'e factor for the spectral lines studied in detail in Sec.~\ref{sec:StkV}}

\tablehead{
\colhead{$\lambda$ [nm]} & \colhead{Transition} & \colhead{Energies [cm$^{-1}$]} & \colhead{A$_{ul}$ [$10^7$ s$^{-1}$]} & \colhead{g$_{\rm eff}$} }

\startdata
259.608 &         b$^4$F$_{9/2}$ - z$^4$H$^{\rm o}_4$   &      22 637 - 61 156  & $0.12$ & 2.5 \\
262.148 & $\quad$ b$^4$F$_{4}$   - z$^4$G$^{\rm o}_4$   &      22 810 - 60 956  & $3.40$ & 1.1 \\
262.119 & $\quad$ a$^6$D$_{3/2}$ - z$^6$D$^{\rm o}_3/2$ & $\quad$ 862 - 39 013  & $0.43$ & 1.9 \\
262.908 & $\quad$ a$^6$D$_{1/2}$ - z$^6$D$^{\rm o}_3/2$ & $\quad$ 977 - 39 013  & $8.70$ & 1.5 \\
\enddata
\end{deluxetable*}

To study in detail the actual applicability of the WFA to infer the longitudinal magnetic field, we select four
spectral lines, relatively close in wavelength, and infer the magnetic field from a theoretical circular polarization profile
calculated in the FAL-C model imposing a certain magnetic field stratification.
Table~\ref{tab:zeeman} summarizes the atomic data of the four chosen transitions, also
highlighted in green in the Grotrian diagram
of Fig.~\ref{fig:grotrian}.
Among these spectral lines we 
find two resonant transitions (262.119 and  262.908~nm) and two 
transitions with metastable lower levels in the b$^4$F term (259.608 and 262.148~nm).

\begin{figure*}[ht!]
\epsscale{1.}
\plotone{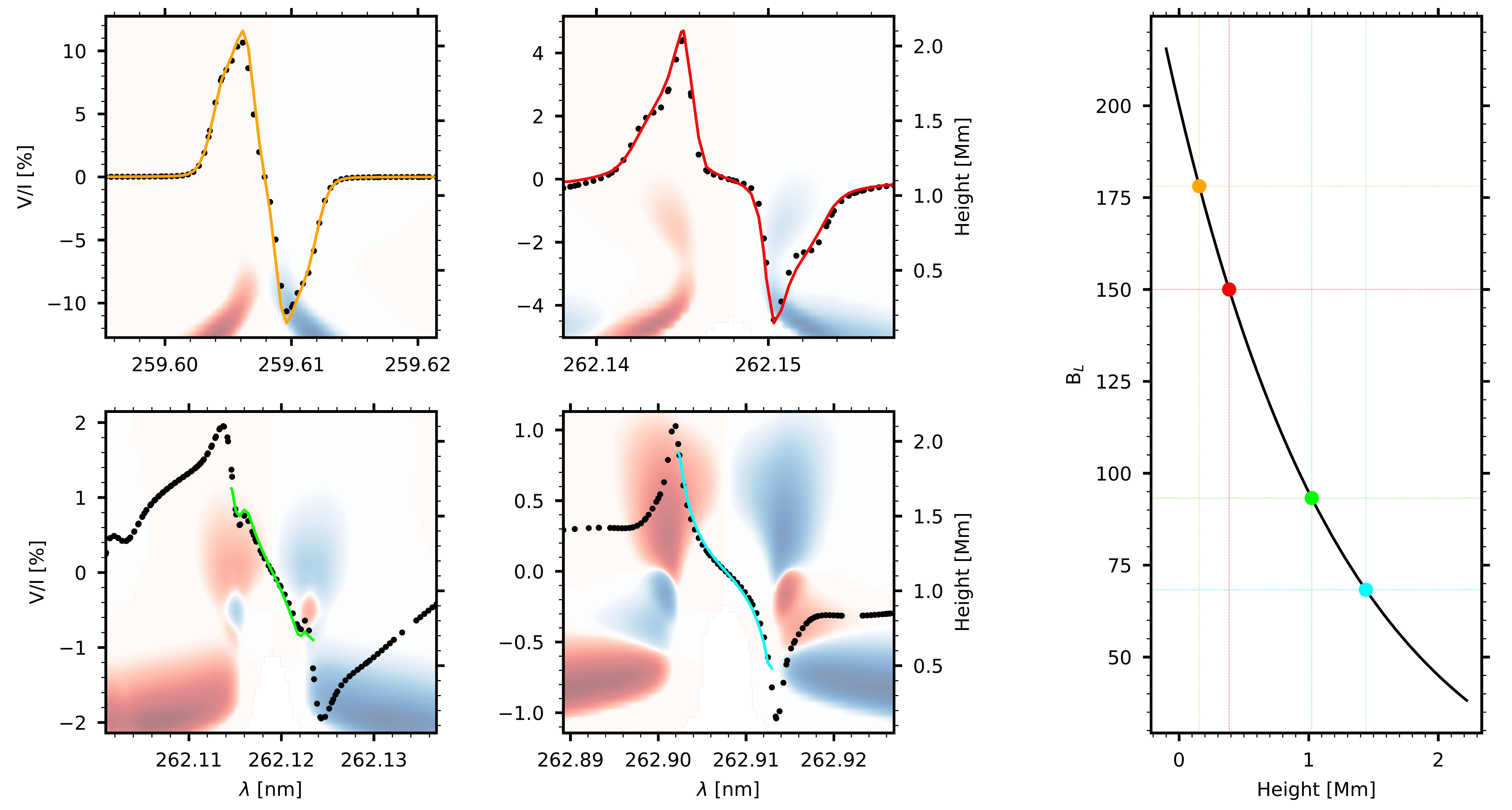}
\caption{Emergent fractional circular polarization
$V/I$ (first and second columns)
for the four selected lines
(see Tab.~\ref{tab:zeeman})
for a LOS at $\mu=1$, calculated in the FAL-C model with a vertical magnetic field with an exponential
stratification (see rightmost panel).
The colored background shows the logarithm of the normalized contribution function for $V/I$ (with the red and blue
colors indicating positive and negative contributions, respectively).
The colored curves in the first and second columns show the fit with the WFA. The colored circles
in the right panel show the $B_L$ value from the fit of the circular polarization profile for the curve
of the same color, plotted at its intersection with the $B_L$ stratification of the model. The colored
lines in the right panel help to quickly identify the magnetic field and height values for each marker.
\label{fig:wfaFe}}
\end{figure*}

The dotted curves in the first and second column in Fig.~\ref{fig:wfaFe} show the fractional circular
polarization profile of the radiation emerging from the FAL-C 
model in which we have added a longitudinal magnetic field
decreasing exponentially with height, from 200~G at the base of the photosphere to 40~G
at the top of the chromosphere.
The colored background in the same panels show 
the logarithm of the normalized contribution function for Stokes $V$, which gives
an idea of the regions in the model atmosphere which contribute 
the most to the emergent circular polarization profile
at each wavelength.
The main contribution to the circular polarization of the spectral line at 259.608~nm comes from the
photosphere, below the temperature minimum. 
The line at 262.147~nm forms at slightly larger heights, with significant contribution from the regions
just above the temperature minimum.
Finally, the contribution function of the 
resonance lines at 262.119 and 262.908~nm have their 
peak in the upper layers of the chromosphere, mainly 
around 1~Mm for the former, and even higher for the latter.
The colored curves show the WFA fit to the theoretical circular polarization profiles.
Because the contribution function of the 259.608 and 262.148~nm lines is more or less concentrated
and coherent in sign at every wavelength, it is possible to find a good fit to the whole spectral range.
The profiles are fitted with a magnetic field of 178 and 150~G, respectively. In the model atmosphere,
these fields are found
at heights $\sim0.2$ and $\sim0.4$~Mm, which correspond to
approximately the middle of the region in the model with significant contribution function values (see colored
background in the corresponding panels).
For the resonant lines at 262.119 and 262.908~nm, the contribution function is significantly more
extended and it changes sign with height at some wavelengths. For this reason, it is not possible to find
a unique satisfactory fit to the whole spectral range (different regions of the profile are being affected
by different magnetic fields, because they do not form 
in the same atmospheric region). We can fit the central
part of the profiles with a magnetic field of 93 and 68~G for the two lines, respectively. These magnetic
fields are found in the model atmosphere at heights $\sim1$ and $\sim1.4$~Mm, respectively, which also
correspond to approximately those heights showing the largest value in the contribution function at the
fitted wavelengths (see colored background in the corresponding panels). The WFA thus seem suitable for
the quick inference of the longitudinal magnetic field at different layers of the solar atmosphere by
combining the diagnostic of multiple \ion{Fe}{2} with different formation regions.

\section{Summary and conclusions} \label{sec:conclusions}

We have theoretically investigated the formation of the intensity and polarization of the    
\ion{Fe}{2} spectral lines in the relatively unexplored 250~-~278~nm window of the near-UV solar spectrum.
We identified those lines for which we predict significant linear or
circular polarization signals and investigated their magnetic sensitivity, as well as their suitability
to infer the magnetic field in the solar atmosphere.

By means of NLTE RT calculations in a semi-empirical model 
of the solar atmosphere, we have obtained the 
Stokes profiles of the emergent spectral line radiation. To this end, we built
a model atom with all possible transitions of interest in the above mentioned spectral region. Due to the
lack of solar observations in this region of the solar spectrum, we use the observed intensity flux
of $\alpha$ Cen A (a solar-like star) in order to adjust one of 
the most important and approximated parameters in our atomic
model, namely the rate of inelastic collisions with electrons. We find that by reducing all these rates by
a factor ten we obtain a theoretical flux very similar to the $\alpha$ Cen A observation at the core of the
\ion{Fe}{2} resonant lines. It is important to emphasize that this region of the spectrum is severely
affected by the so-called UV blanketing, which we cannot accurately include in our modeling, and that is
evident when comparing the continuum level between the theoretical and observed flux profiles. While the
impact of the UV blanketing, a priori, limits the diagnostic capabilities of the weaker lines and the
wings of the strongest lines, the metallicity of $\alpha$ Cen A is considerably larger than that of the Sun.
Therefore the impact of UV blanketing is expected to be
less significant in the solar spectrum than 
in $\alpha$ Cen A. Future solar observations
in this spectral window will
 make possible to study in detail how the
 UV blanketing affects the conclusions of
this work and to further validate and improve
the atomic data for the modeling of the \ion{Fe}{2} spectral lines.

Most of the stronger (in intensity) \ion{Fe}{2} lines do not
show remarkable linear polarization signals for a close to the limb LOS ($\mu=0.1$). Nevertheless, there
are numerous lines with predicted signals with amplitudes larger 
than $0.2$~\%, and therefore we expect that they 
can be observed with spectropolarimetric instrumentation similar to that of CLASP2.
We show the region of formation,
critical Hanle field, and information about blends for all these lines. Most of them form around the middle
chromosphere, with some forming in the upper chromosphere. We study in detail the magnetic sensitivity of
the linear polarization of five of these lines, showing that those forming in the middle (upper) chromosphere
are typically sensitive to magnetic fields with strength up to $\sim50$~G ($\sim20$~G). These lines thus
encode information about the expected weak magnetic fields in the chromosphere of the quiet Sun.

We find that, for a magnetic field strength of $50$~G, more than a hundred \ion{Fe}{2} transitions
are predicted to show circular polarization signals above $0.5$\%\, strong enough to be able of use the WFA to reliably infer the longitudinal component of the magnetic field. We show the region of formation,
effective Land\'e factor, and information about blends for all these lines. Most of them form in the upper
photosphere and lower chromosphere, with some of them forming in the upper chromosphere. 
We estimate the
order of magnitude of the magnetic field for which the WFA may not hold. 
We find that, for the lines forming in the photosphere, these values are mostly between 200 and
400~G, while in the middle and upper chromosphere the typical values are between 700 and 900~G.
By applying the WFA to the theoretical circular polarization profiles of four lines with different
heights of formation we are able to recover 
the magnetic field values of the imposed exponential stratification.
When checking the height in the atmospheric model at which the inferred magnetic field is located, we can
see that they coincide with the information provided by the circular polarization contribution function
regarding the region of formation of each of the selected spectral lines.

The narrower 279.2~-~280.7~nm spectral window of the CLASP2 suborbital space experiment includes several spectral lines of already demonstrated utility for diagnosing the magnetic field across the whole solar chromosphere, namely the \ion{Mg}{2} h and k lines and the resonance lines of \ion{Mn}{1}. In addition, as we shall show in a forthcoming publication, it includes several other weaker spectral lines with measurable circular polarization signals in active region plages, such as two hitherto unexplored \ion{Fe}{2} lines. The circular polarization of these \ion{Fe}{2} lines provides information about the magnetic field which is complementary to the \ion{Mg}{2} and \ion{Mn}{1} lines in terms of the region of the atmosphere that they sample. All these  near-UV lines of the CLASP2 spectral region are very suitable to map the magnetic field from the lower to the upper chromosphere.
Nevertheless, the results of the present investigation on the magnetic sensitivity of the \ion{Fe}{2} lines in the 250~-~278~nm spectral region, especially the prediction that they should show measurable circular polarization signals and sample different layers of the solar chromosphere when suitably combined, lead us to conclude that including these lines would further help determine the magnetic field throughout the whole solar chromosphere.

\acknowledgements

We acknowledge the funding received from the European Research Council (ERC)
under the European Union's Horizon 2020 research and innovation programme (ERC
Advanced Grant agreement No 742265).

\bibliography{feii}
\bibliographystyle{aasjournal}

\end{document}